\documentclass[12pt]{article}

\usepackage{graphicx}
\usepackage{setspace}
\usepackage{scicite}
\usepackage{upgreek}
\usepackage{bm}

\usepackage{times}

\newenvironment{sciabstract}{%
\begin{quote} \bf}
{\end{quote}}

\newcounter{lastnote}
\newenvironment{scilastnote}{%
\setcounter{lastnote}{\value{enumiv}}%
\addtocounter{lastnote}{+1}%
\begin{list}%
{\arabic{lastnote}.}
{\setlength{\leftmargin}{.22in}}
{\setlength{\labelsep}{.5em}}}
{\end{list}}

\title{[C\,\textsc{ii}] 158$\upmu$m Emission from the Host Galaxies of Damped Lyman Alpha Systems}

\author
{Marcel Neeleman$^{1\ast}$, Nissim Kanekar$^{2}$, J. Xavier Prochaska$^{1}$, Marc Rafelski$^{3}$,\\
Chris L. Carilli$^{4,5}$, Arthur M. Wolfe$^{6 \dagger}$\\
\\
\normalsize{$^{1}$University of California Observatories-Lick Observatory,}\\
\normalsize{University of California, Santa Cruz, CA, 95064, USA}\\
\normalsize{$^{2}$National Centre for Radio Astrophysics, Tata Institute of Fundamental Research,}\\
\normalsize{Pune University, Pune 411007, India}\\
\normalsize{$^{3}$Space Telescope Science Institute, Baltimore, MD 21218, USA}\\
\normalsize{$^{4}$National Radio Astronomy Observatory, Socorro, NM 87801, USA}\\
\normalsize{$^{5}$Cavendish Laboratory, University of Cambridge,}\\
\normalsize{19 J. J. Thomson Ave., Cambridge CB3 0HE, UK}\\
\normalsize{$^{6}$Department of Physics and Center for Astrophysics and Space Sciences,}\\
\normalsize{University of California, San Diego, CA, 95064, USA}\\
\normalsize{$^{\dagger}$Deceased}
\\
\normalsize{$^\ast$Corresponding author. E-mail: marcel@ucsc.edu.}
}

\date{}

\begin{document} 


\baselineskip24pt


\maketitle


\begin{sciabstract}
Gas surrounding high redshift galaxies has been studied through 
observations of absorption line systems toward background quasars for decades. 
However, it has proven difficult to identify and characterize the galaxies associated
with these absorbers due to the intrinsic faintness of the galaxies compared to 
the quasars at optical wavelengths. Utilizing the Atacama Large Millimeter/Submillimeter 
Array, we report on detections of [C\,\textsc{ii}]~158$\bm{\upmu}$m line and dust continuum 
emission from two galaxies associated with two such absorbers at a redshift of 
$\bm{z \sim 4}$. Our results indicate that the hosts of these high-metallicity absorbers 
have physical properties similar to massive star-forming galaxies and are 
embedded in enriched neutral hydrogen gas reservoirs that extend well beyond the 
star-forming interstellar medium of these galaxies.
\end{sciabstract}

Gas plays a crucial role in the formation of galaxies by providing the fuel for star 
formation. However, the initial overdensities of baryons and dark matter from which 
galaxies form do not contain enough gas to sustain the observed star formation rate 
(SFR) of galaxies \cite{Prochaska2005,Keres2005}. Galaxies must therefore accrete 
gas from their surroundings. A comprehensive understanding of the 
accreting gas, in particular neutral hydrogen (H\,\textsc{i}) gas, which is a crucial 
component \cite{Keres2005}, is thus critical for understanding the 
formation and evolution of galaxies. 
Unfortunately, observing H\,\textsc{i} gas in emission from galaxies at 
cosmological distances is challenging because of the intrinsic weakness of its most 
reliable tracer, the H\,\textsc{i}~21~cm hyperfine structure line. This line is difficult to detect 
at even moderate redshifts with today's telescopes \cite{Fernandez2016}, and
at high redshifts, $z \geq 1$, we must rely on studying H\,\textsc{i} gas in absorption.

As light from a bright background source, such as a quasar, travels toward us, 
it encounters pockets of H\,\textsc{i} gas which imprint a characteristic absorption signature 
in the spectrum due to absorption at the frequency of the redshifted Lyman-alpha (Ly-$\upalpha$) line. 
The strongest H\,\textsc{i} absorption features in quasar spectra have H\,\textsc{i} column 
densities $\geq 2 \times 10^{20}$~cm$^{-2}$, and are known as 
damped Ly-$\upalpha$ Absorbers (DLAs) because their Ly-$\upalpha$ 
absorption line profiles show distinctive damping wings \cite{Wolfe2005}. 
Observations of DLAs have revealed a wealth of information on the properties 
of the absorbing gas, including its kinematical signatures \cite{Prochaska1997}, metal 
enrichment \cite{Rafelski2012}, dust content \cite{Pettini1994} and 
temperature \cite{Kanekar2014}. 
Moreover, DLAs have been intimately linked to galaxies through direct imaging 
studies at intermediate redshifts \cite{Moller2002,Krogager2012}, 
scaling relations \cite{Neeleman2013,Christensen2014}, and cosmological simulations 
\cite{Cen2012,Bird2014}. Surveys of DLAs therefore provide a powerful means to 
study H\,\textsc{i} gas surrounding high redshift galaxies.

Unfortunately, directly detecting the starlight from the foreground galaxy that is associated 
with the DLA is challenging at optical wavelengths, owing to the presence of the much brighter 
background quasar \cite{Fynbo2010}. As a 
result, despite many recent searches \cite{Peroux2011,Krogager2012,Fumagalli2015,Srianand2016}, 
there are only a handful of high-redshift DLAs whose host galaxies have been identified by 
imaging studies. As such, there is little direct observational data on the nature of the 
galaxies that give rise to DLAs at different redshifts, including their mass, SFR, and the 
extent of the neutral H\,\textsc{i} gas surrounding the galaxy. At the same time, extensive 
data is available from deep optical and near-infrared surveys on the stellar properties of 
luminosity-selected galaxies \cite{Giavalisco2004,Beckwith2006,Grogin2011}, 
but little is known about the gas surrounding these galaxies. A crucial link is therefore 
missing in connecting high redshift galaxies with the gaseous envelopes that shape their 
evolution.

We report here on a search for the galaxies associated with two DLAs at $z \sim 4$ using
the Atacama Large Millimeter/Sub-millimeter Array (ALMA). At sub-millimeter wavelengths, the
background quasars emit little radiation, enabling a search for the host galaxy of the DLA 
at close angular separation (impact parameter) from the quasar. The DLAs towards quasars 
SDSS~J081740.52+135134.5 and SDSS~J120110.31+211758.5 were selected from a large 
sample of DLAs because of their higher than average metal content \cite{Rafelski2012}. 
We used ALMA to carry out a search for emission in both the singly-ionized 
carbon fine-structure line at a rest wavelength of 157.74~$\upmu$m 
([C\,\textsc{ii}]~158$\upmu$m) and the far-infrared (FIR) dust continuum 
from the two DLA host galaxies \cite{method}. The [C\,\textsc{ii}]~158$\upmu$m line is 
expected to be the strongest FIR line from galaxies at these redshifts \cite{Carilli2013},
in part because it is the primary coolant of cold H\,\textsc{i} gas \cite{Pineda2013}, and 
because it is the strongest observed FIR line in the local Universe \cite{Malhotra2001}. 

In each of the ALMA observations, we detect a $>$6$\sigma$ emission feature 
at the frequency of the redshifted [C\,\textsc{ii}]~158$\upmu$m line at $z=4.2601$ and 
$z=3.7978$, which is offset from the quasar position by 6.2$''$ and 2.5$''$, for
SDSS~J081740.52+135134.5 and SDSS~J120110.31+211758.5, respectively (Fig.~1).
The redshifts of the DLAs, as measured from low ionization atomic absorption lines 
(e.g., singly ionized iron and silicon), and the [C\,\textsc{ii}]~158$\upmu$m emission (Fig.~2)
are within 100~km~s$^{-1}$ of each other (Fig.~2). The close proximity in both redshift
space and angular separation on the sky indicates that the [C\,\textsc{ii}]~158$\upmu$m 
emission must come from a galaxy associated with the DLA.

Our  [C\,\textsc{ii}]~158$\upmu$m images are sensitive to the kinematics of the 
galaxy, as any gas motion will yield a Doppler shift in the observed 
[C\,\textsc{ii}]~158$\upmu$m line frequency. Since the emission is 
only barely resolved spatially, the kinematic signature of the gas is best described by 
a position-velocity ({\it p-v\/}) diagram (Fig.~3). For ALMA~J081740.86+135138.2, 
the shape of the emission is indicative of rotation, which is corroborated by the 
`double-horned' profile seen in the [C\,\textsc{ii}]~158$\upmu$m emission spectrum (Fig.~2).
From the magnitude of the rotation, we estimate a lower limit to the 
dynamical mass of $6 \times 10^{10}$~solar~masses~(M$_\odot$) for this galaxy \cite{method}. 
For ALMA~J120110.26+211756.2, the {\it p-v\/} diagram suggests more complicated 
gas dynamics which is corroborated by the spatial offset between the redshifted 
and blueshifted components \cite{method}.

FIR dust continuum emission co-spatial with the [C\,\textsc{ii}]~158$\upmu$m emission
is also detected in the ALMA observations (Fig~1). The dust continuum yields estimates 
of the SFR, as emission is due to reprocessed starlight, mainly from the youngest stars 
\cite{method}. We find that both galaxies are forming stars at moderately high rates, 
with SFR of $110 \pm 10$ M$_\odot~{\rm yr}^{-1}$ for 
ALMA~J081740.86+135138.2 and $24 \pm 8$~M$_\odot~{\rm yr}^{-1}$ for 
ALMA~J120110.26+211756.2. The total far-infrared (TIR) luminosities of the galaxies are
estimated by fitting a modified black body spectrum to the observed FIR dust continuum
measurement. The properties of the galaxies are tabulated in Table~S2.

The observations enable a comparison between the properties of these
absorption-selected galaxies and the properties of emission-selected galaxies observed at
similar redshifts, such as Lyman-break galaxies (LBGs). LBGs are star-forming galaxies 
identified in photometric studies by their characteristic rest-frame ultraviolet luminosity
deficit. This deficit is due to the absorption of photons blueward of the Lyman 
continuum break \cite{Steidel1996}. The observed SFRs of the two absorption-selected 
galaxies are comparable to the SFR of bright LBGs at similar redshift \cite{Capak2015}. However, 
by selecting the DLAs on their high metallicity, we are probing the massive end of the 
distribution of DLA host galaxies \cite{Neeleman2013,Christensen2014}. 
The typical galaxies associated with DLAs likely have a smaller mass, and thus,
a smaller SFR, consistent with non-detections reported in the literature 
\cite{Fumagalli2015}.

Plotting the [C\,\textsc{ii}]~158$\upmu$m line luminosity versus the SFR for high-redshift 
($z \geq 1$) galaxies (Fig.~4a) shows that our absorption-selected galaxies occupy 
the same parameter space as high-redshift, emission-selected galaxies, and fall within 
one standard deviation of the correlation found in the local Universe between these 
two observables \cite{DeLooze2014}. Similarly, the two absorption-selected galaxies 
fall within the same parameter space as their high-redshift emission-selected counterparts 
in a plot of the ratio of [C\,\textsc{ii}]~158$\upmu$m line luminosity to TIR luminosity 
versus TIR luminosity (Fig.~4b). This strengthens the assertion that high-metallicity, 
absorption-selected galaxies are similar to moderately star-forming, emission-selected 
galaxies, such as the massive end of the LBG population. This assertion is further corroborated 
by the agreement between our dynamical mass estimate in ALMA~J081740.86+135138.2 
and the dynamical mass estimates of comparable emission-selected galaxies at 
$z \sim 5$ \cite{Capak2015}. The observed connection between massive LBGs and 
high-metallicity DLAs supports results from earlier DLA studies at lower 
redshifts \cite{Moller2002,Rafelski2016} and cross-correlation studies with the Ly~$\upalpha$
forest \cite{Font-Ribera2012}.

We can use the impact parameter estimates to probe the physical extent of the H\,\textsc{i} 
gas around absorption-selected galaxies. The observed impact parameters
correspond to physical (proper) distances of 42~kpc and 18~kpc at the DLA 
redshifts for ALMA~J081740.86+135138.2 and ALMA~J120110.26+211756.2, respectively. 
These distances are substabtially larger than the extent of the [C\,\textsc{ii}]~158$\upmu$m line emission, 
which extends to $\sim 5$~kpc, and indicate that these galaxies have large reservoirs of 
H\,\textsc{i} relatively far away from their star-forming regions. The observed distances also 
require that the emission and absorption lines arise in physically separated gas. Therefore, the absorbing gas
must probe either H\,\textsc{i} gas associated with a satellite galaxy, 
an enriched neutral outflow from the galaxy, or H\,\textsc{i} gas in the inner circumgalactic 
medium/extended disk of the galaxy. 
The first explanation is disfavored, because the satellite galaxies would need to be 
metal-enriched and show highly turbulent velocity dispersions of several hundreds of 
km~s$^{-1}$, indicative of massive star-forming systems. However, little 
star-formation is observed, as no [C\,\textsc{ii}]~158$\upmu$m emission is seen at 
the position of the absorber.

The absorbing gas is therefore more likely to reside in the inner gaseous halo of 
each DLA host galaxy. For the DLA towards SDSS~J081740.52+135134.5, the gas is 
systematically blueshifted along the same direction and with the same magnitude as the 
rotation of the cool gas disk observed in [C\,\textsc{ii}]~158$\upmu$m line emission. 
This could indicate that the gas detected in absorption is co-rotating in an extended disk.
Most simulations predict precisely such an extended planar configuration fed by 
cold flows \cite{Stewart2013} with properties similar 
to those observed in the $z = 4.26$ DLA \cite{Danovich2015}. The DLA 
towards SDSS~J120110.31+211758.5 is harder to classify because its absorption is seen 
over the full velocity range of the emission profile, which spans almost 500~km~s$^{-1}$. 
This is more indicative of a large-scale outflow or a highly perturbed system, which is 
corroborated by the spatial shift in the [C\,\textsc{ii}]~158$\upmu$m line emission (Fig.~S2).
In both cases, it is clear that the DLA host galaxy has effectively enriched its inner gaseous 
halo, in agreement with recent simulations \cite{Muratov2015}. 

Our results indicate that the galaxies giving rise to high-metallicity DLAs have characteristics 
similar to the high mass end of the LBG population \cite{Moller2002,Font-Ribera2012}, 
and are embedded in a large reservoir of neutral H\,\textsc{i} gas. This gas is being 
enriched by the galaxy, but is bound to it, as there is almost no systematic velocity offset 
between the metals seen in absorption and the [C\,\textsc{ii}]~158$\upmu$m line in emission. 
This suggests that this halo gas will eventually accrete back onto the galaxy providing 
enriched gas for future star formation.

\begin{scilastnote}
\item Support for this work was provided by the NSF through award SOSPA2-002 from the NRAO. 
MN and JXP are partially supported by a grant from the National Science Foundation (AST-1412981).
NK acknowledges support from the Department of Science and Technology via a 
Swarnajayanti Fellowship (DST/SJF/PSA-01/2012-13), and MR was partially supported by a 
NASA Postdoctoral Program fellowship. ALMA is a partnership of ESO (representing 
its member states), NSF (USA) and NINS (Japan), together with NRC (Canada), NSC and 
ASIAA (Taiwan), and KASI (Republic of Korea), in cooperation with the Republic of Chile. 
The Joint ALMA Observatory is operated by ESO, AUI/NRAO and NAOJ. Part of the data 
presented herein were obtained at the W.M. Keck Observatory, which is operated as a 
scientific partnership among the California Institute of Technology, the University of California 
and the National Aeronautics and Space Administration. The Observatory was made possible 
by the generous financial support of the W.M. Keck Foundation. The authors wish to recognize 
and acknowledge the very significant cultural role and reverence that the summit of Mauna Kea 
has always had within the indigenous Hawaiian community.  We are most fortunate to have the 
opportunity to conduct observations from this mountain. Finally we wish to acknowledge the Esalen
Institute, whose natural beauty was the catalyst for writing this paper. The data reported in this paper 
are available though the ALMA archive (https://almascience.nrao.edu/alma-data/archive) with
project code: ADS/JAO.ALMA\#2015.1.01564.S, and the Keck Observatory Archive 
(https://koa.ipac.caltech.edu) with program ID: U163HR.
\end{scilastnote}

{\noindent \bf Supporting Material}\\
Supporting text\\
Figures S1 -S2\\
Tables S1 - S4\\
References \cite{ALMA2016,McMullin2007,Greisen2003,Vogt1994,Asplund2009,Solomon1997,Goldsmith2012,Calzetti2010,LonsdalePersson1987,Wang2013,Neeleman2016,Abazajian2009}

\clearpage

\begin{figure}[!h]
\centering
\includegraphics[width=0.71\textwidth]{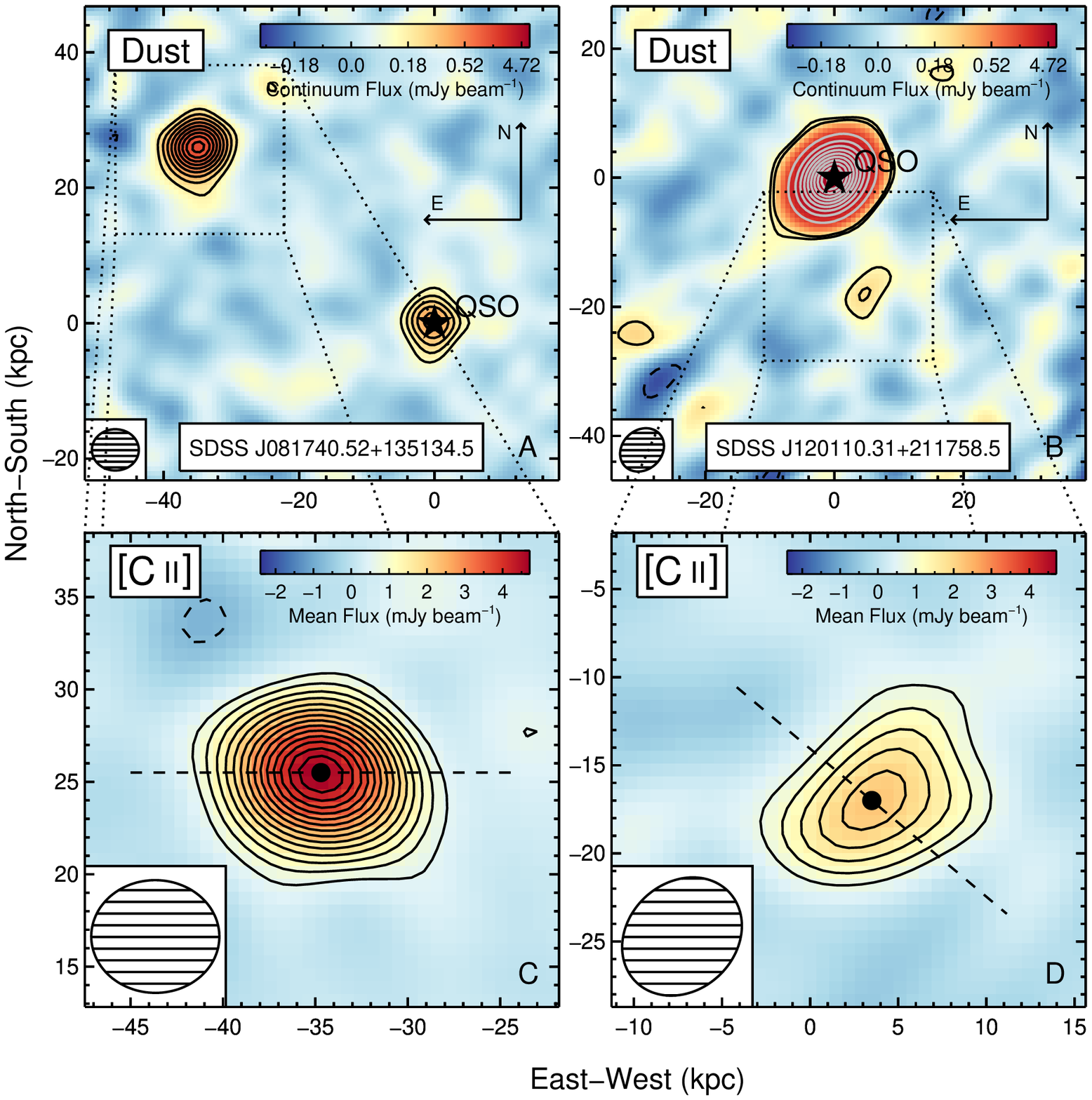}
\end{figure}
\noindent {\bf Fig. 1. 400~GHz continuum and [C\,\textsc{ii}]~158$\bm{\upmu}$m emission from two DLA fields.}
Panels A and B show the $\approx 400$~GHz continuum emission from the regions 
surrounding two quasars (black stars). Black contours begin
at 3$\sigma$ and increase by $\sqrt{2}\sigma$; dashed contours indicate negative values. Gray 
contours are drawn at increments of 25$\sigma$. The axes give the relative physical (proper) distance 
at the DLA redshifts (i.e., $z=4.2584$ and $z = 3.7975$ for SDSS~J081740.52$+$135134.5 and 
SDSS~J120110.31$+$211758.5, respectively). Panels C and D show the mean flux density over the
full [C\,\textsc{ii}]~158$\upmu$m line profile displayed in panels A and B of Fig.~2 for a smaller 
region centered on the identified DLA host galaxies. No other emission lines are detected in these
fields. The line contours begin at 3$\sigma$, with each subsequent contour 
increasing by $\sqrt{2}\sigma$. The size of the synthesized beam is shown in the bottom left of 
each panel. The dashed line is the measured major axis of the galaxy QSO, quasar or quasi-stellar object
\cite{method}.

\newpage

\begin{figure}[!h]
\centering
\includegraphics[width=\textwidth]{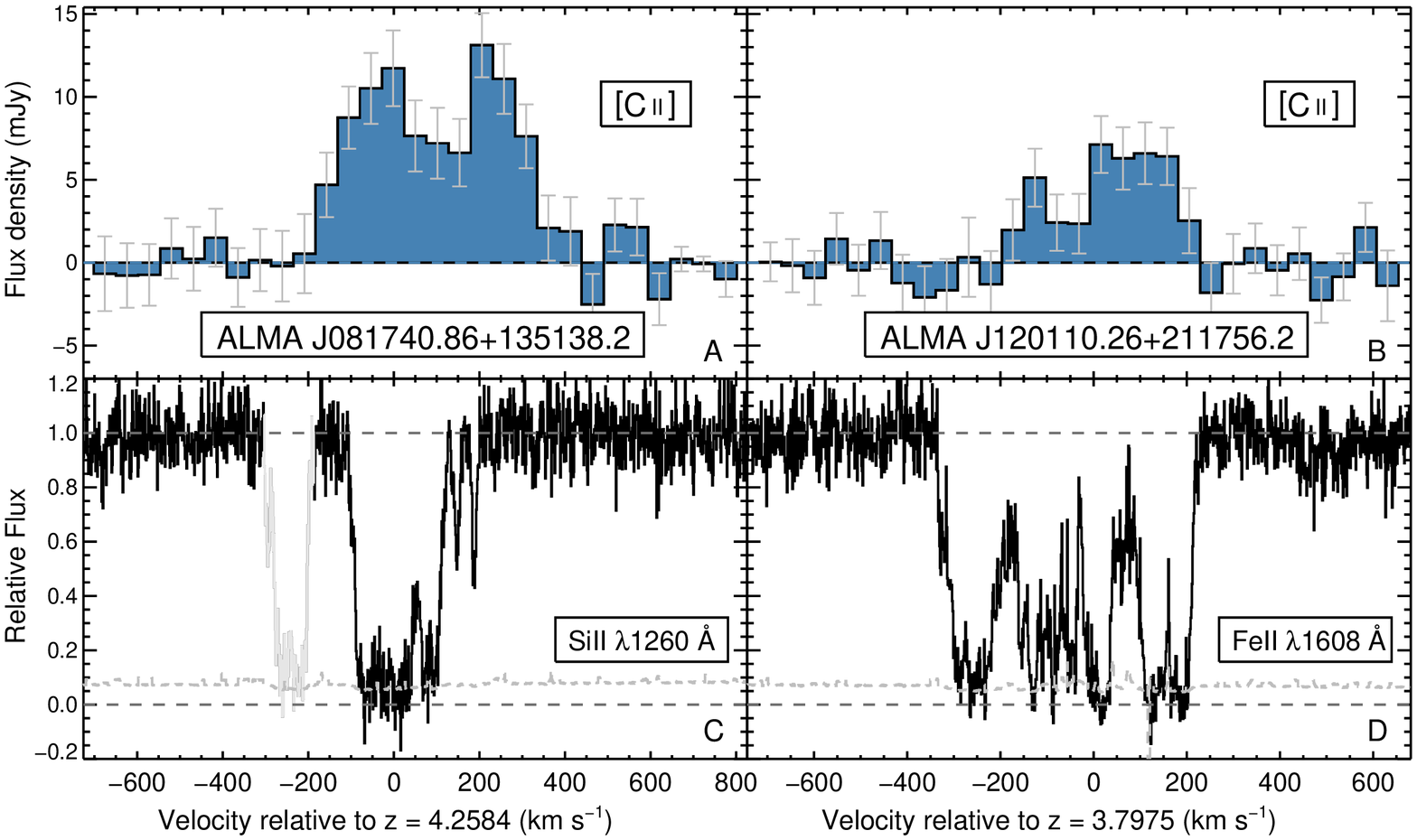}
\end{figure}
\noindent {\bf Fig.~2. Emission and absorption spectra from the host galaxies and DLAs.}
Panels A and B show the [C\,\textsc{ii}]~158$\upmu$m emission profile for the galaxy hosts
identified with two high redshift DLAs (Fig.~1). The 1$\sigma$ uncertainties are shown in gray.
The zero-point of the velocity scale was chosen to 
correspond with the strongest absorption feature of the DLAs (panels C and D). 
The absorption profiles are for two representative low-ionization elements, singly ionized silicon 
and singly ionized iron, which trace the bulk of the metals in the absorbers. The grayed-out 
region in panel C is a sulfur absorption line (S\,\textsc{ii}~$\uplambda$1259\,{\AA}). The agreement
in redshift and width of the absorption and emission lines indicate that the [C\,\textsc{ii}]~158$\upmu$m
emission is from the DLA host galaxy.

\newpage

\begin{figure}[!h]
\centering
\includegraphics[width=\textwidth]{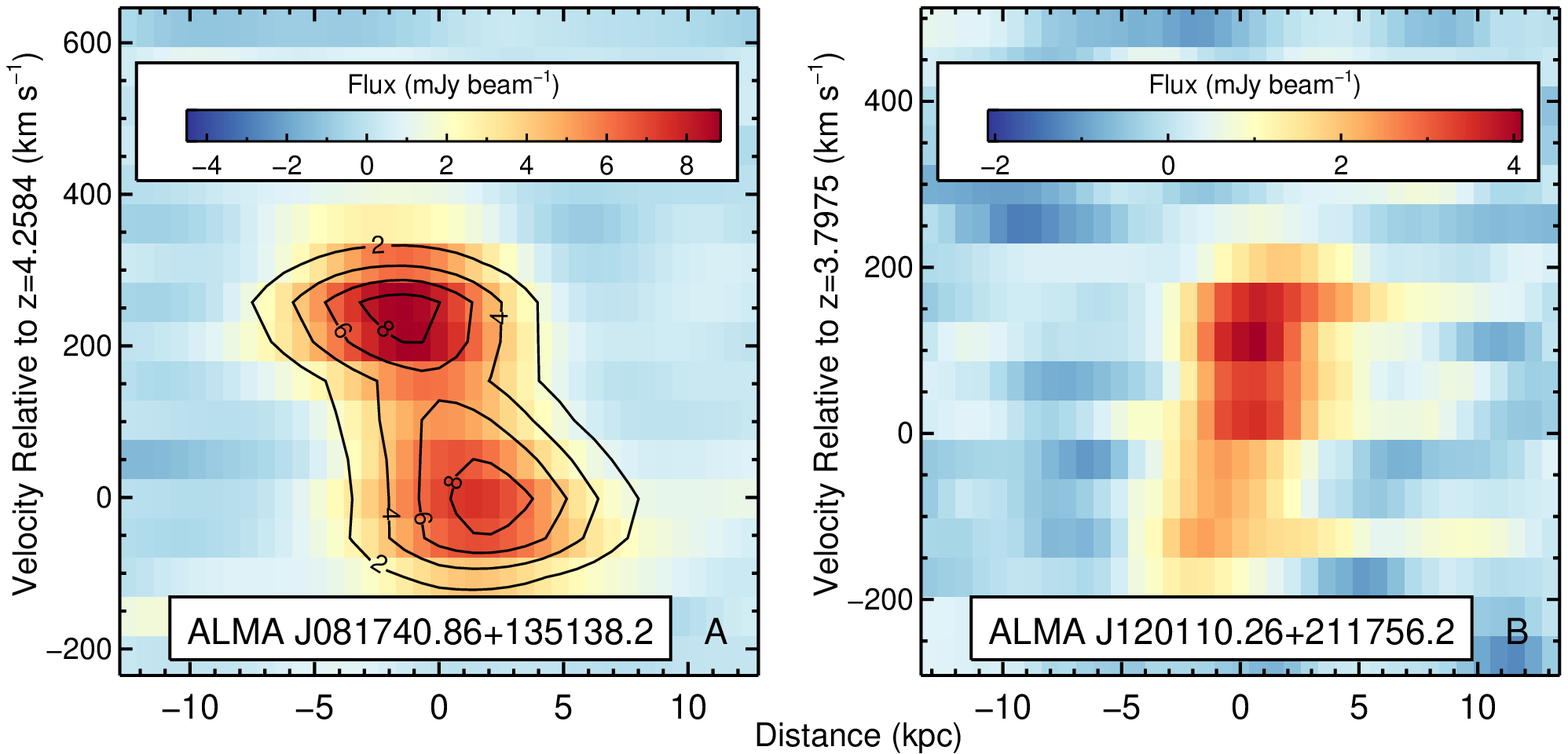}
\end{figure}
\noindent {\bf Fig.~3. Position-velocity (\ dtextbf{\textit{p-v\/}}) diagrams for the [C\,\textsc{ii}]~158$\bm{\upmu}$m emission.} 
The distance is calculated from the center of the [C\,\textsc{ii}]~158$\upmu$m emission 
along the observed major axis of the galaxy (Fig.~1). For ALMA~J081740.86$+$135138.2,
a model of a simple uniformly-rotating disk is shown in contours \cite{method}.  The agreement 
between this model and the data suggest that the [C\,\textsc{ii}]~158$\upmu$m line originates 
from a cool gaseous disk. For ALMA J120110.26$+$211756.2, the [C\,\textsc{ii}]~158$\upmu$m 
emission appears more complicated, albeit at a lower signal-to-noise ratio. mJy, milijansky.

\newpage

\begin{figure}[!h]
\centering
\includegraphics[width=0.496\textwidth]{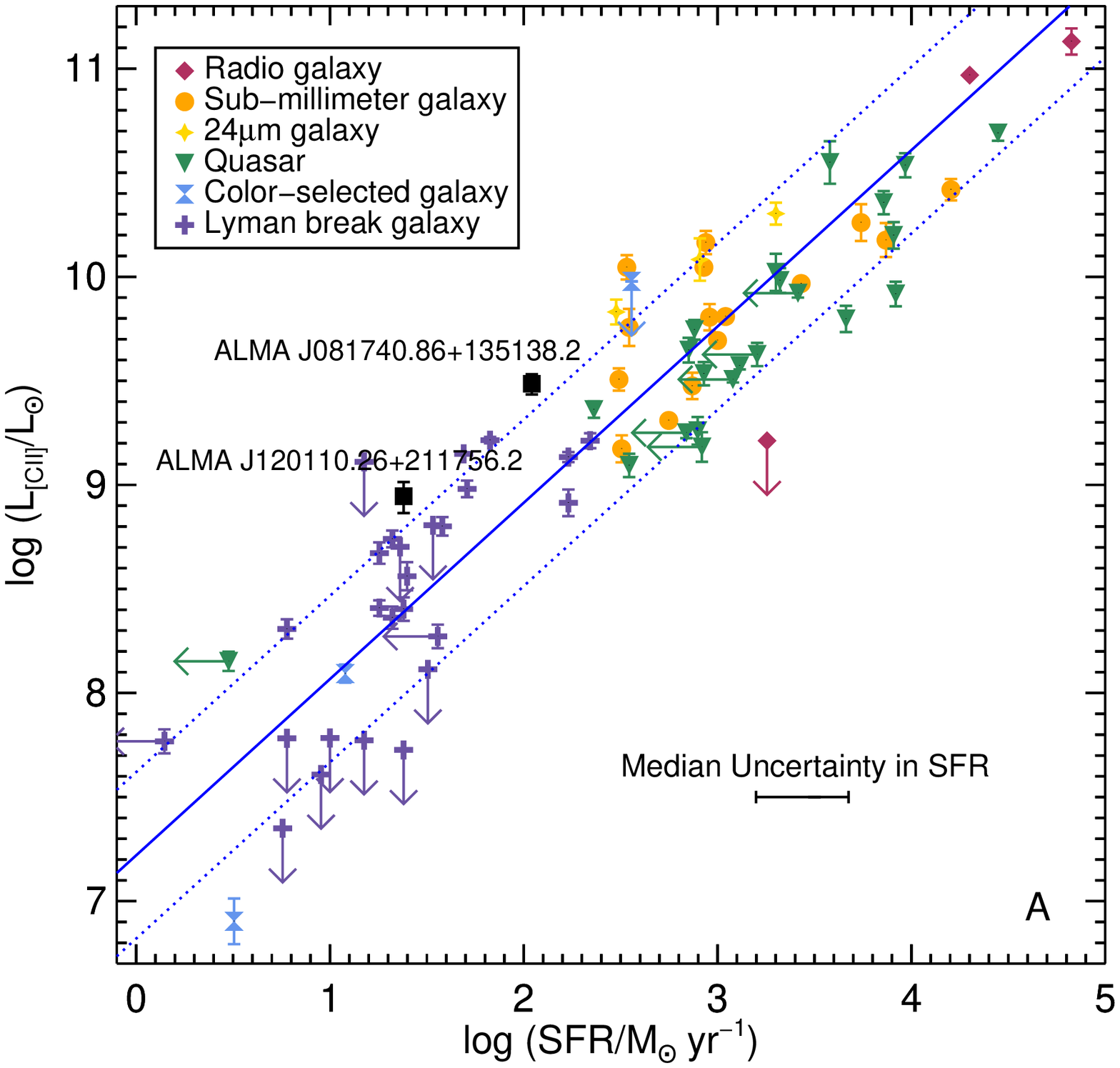}
\includegraphics[width=0.496\textwidth]{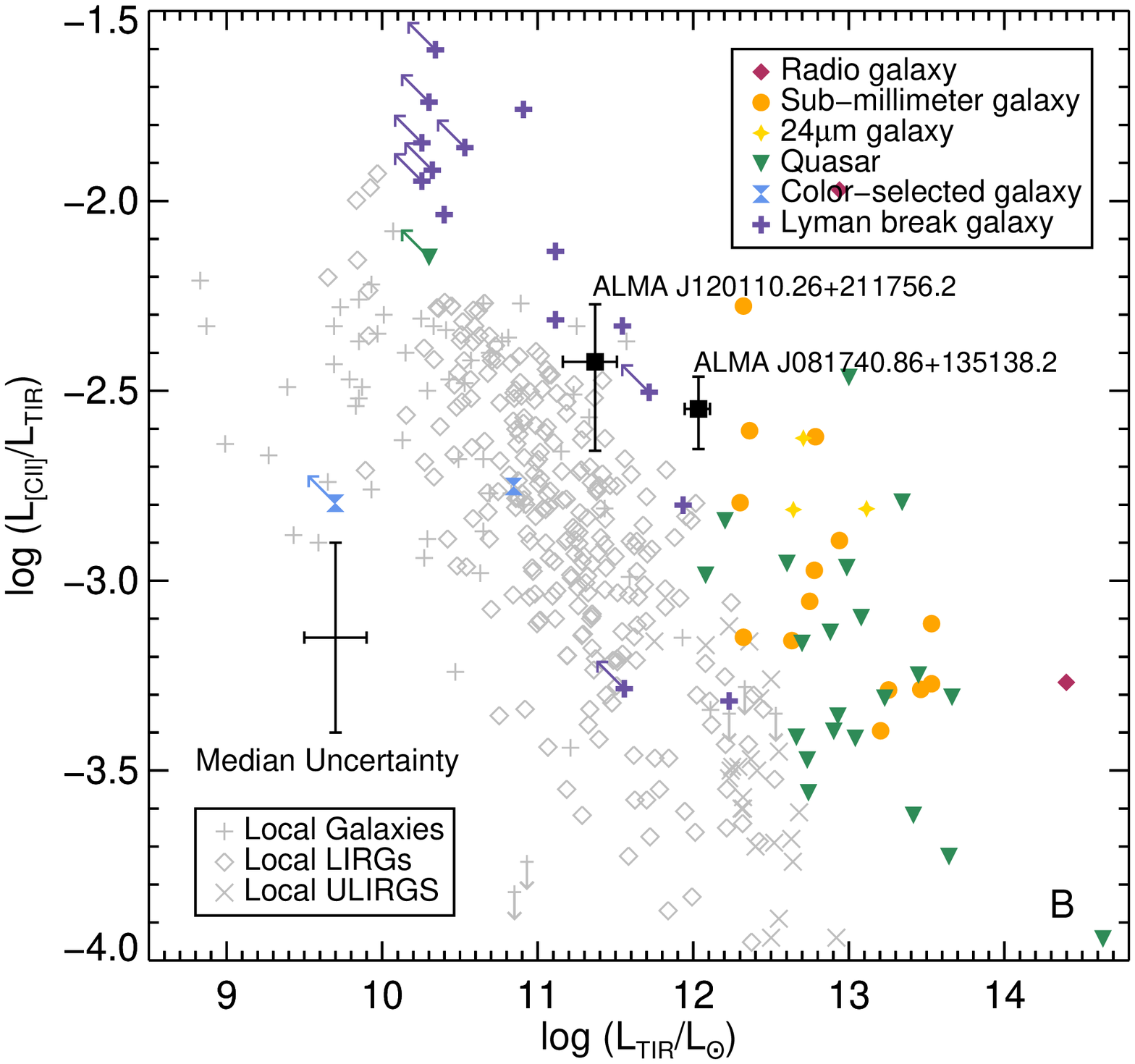}
\end{figure}
\noindent {\bf Fig. 4. Comparison of observed properties of the DLA hosts with local and high 
redshift galaxies.} Panel A shows the correlation between the [C\,\textsc{ii}]~158$\upmu$m 
luminosity and star formation rate (SFR) for a sample of $z > 1$ galaxies categorized by 
galaxy type \cite{Carilli2013}. Also plotted is the correlation for galaxies in the local 
Universe (blue line), along with the associated scatter (blue dashed lines) \cite{DeLooze2014}. 
Panel B shows the ratio of [C\,\textsc{ii}]~158$\upmu$m luminosity to total infrared (TIR)
luminosity, $L_{\rm TIR}$ as a function of $L_{\rm TIR}$ for the same sample of $z > 1$ galaxies. Plotted in gray 
are a sample of local galaxies and local (ultra) luminous infrared galaxies, (U)LIRGs. The 
high-metallicity, absorption-selected galaxies occupy the same parameter space as 
moderately star-forming, emission-selected galaxies at similar redshift, suggesting that they 
have similar characteristics. $L_{\odot}$, luminosity of the Sun. Error bars indicate 1$\sigma$.

\newpage

%
%
%
%
%
%
%
%

\noindent\textbf{Materials and Methods}

\noindent\underline{Cosmology}

Throughout this paper we use a flat $\Lambda$ cold dark matter cosmology, defined by the 
parameters, $\Omega_{\Lambda}=0.7$, $\Omega_{\rm M}=0.3$ and $H_0 = 70$~km~s$^{-1}$.
Adopting this cosmology facilitates comparison with previous results in the literature.\\

\noindent\underline{Observations, Data Reduction and Analysis}

ALMA ({\it 34\/}) observed the two fields surrounding quasars SDSS~J081740.52+135134.5 and
SDSS~J120110.31+211758.5 over a span of four days from 2015~December~29 to 
2016~January~1 (all dates are given in Universal Time). Total on-source integration
times on the two quasar fields were 46~minutes and 76~minutes, respectively. One of the 
ALMA 1.875~GHz bands was centered on the expected redshifted frequency of the 
[C\,\textsc{ii}]~158$\upmu$m line, with the expected [C\,\textsc{ii}]~158$\upmu$m redshift 
determined from the strongest metal line absorption feature seen in the QSO absorption 
spectrum (Fig. 3). The remaining three 2~GHz bands were set to measure the 
continuum emission in the quasar fields. Exact details of the instrument configuration are 
given in Table~S1. A third quasar, SDSS~J144331.17+272436.7, was also observed in May 2016 as part
of the same program. However, problems with quality assurance meant that the
data for this source were not delivered by the observatory until after this paper was submitted;
we therefore do not include it in our study.

The data were calibrated as part of the standard reduction procedure by the ALMA support 
staff ({\it 34\/}). These calibrated visibility data sets were then re-examined using 
both the Common Astronomy Software Application (CASA) package ({\it 35\/}) 
and the Astronomical Image Processing System (AIPS) package ({\it 36\/})
to perform additional flagging of sub-standard channels. The visibility sets from the 
three bands set up for continuum measurements were imaged, using natural weighting, 
yielding synthesized  beams of  $1''\times0.9''$ (SDSS~J081740.52+135134.5)  and 
$1.1''\times0.9''$ (SDSS~J120110.31+211758.5). For SDSS~J120110.31+211758.5, 
we self-calibrated the data set on the quasar continuum emission, resulting in an 
improvement of $\approx 25$\% in the root mean square noise of the final image. The
visibility data set from the band centered on the redshifted [C\,\textsc{ii}]~158$\upmu$m
emission line was imaged, after subtracting any continuum flux, using natural weighting 
to create a spectral data cube with the same resolution as the continuum image. 
The spectral cubes were then averaged by four channels to a resolution of 62.5~MHz, 
corresponding to velocity resolutions of 52~km~s$^{-1}$ and 47~km~s$^{-1}$ for 
SDSS~J081740.52+135134.5 and SDSS~J120110.31+211758.5, respectively. 
After applying the primary beam correction, the final spectral cubes and continuum 
images were analyzed with the imaging functions provided in CASA. The observed 
properties of the continuum image and spectral data cube for the two detected 
galaxies associated with the two DLAs are listed in Table~S2.

The optical spectra shown in Fig.~2 were obtained with the High Resolution (HIRES) 
Spectrograph on the Keck-I telescope ({\it 37\/}). The HIRES spectra were obtained 
on 2011 January 16, 2011 January 24 and 2011 January 25, as part of a program aimed 
at probing the metallicity evolution of the Universe over cosmic time ({\it 6\/}). 
The total Keck-I integration times were 3~hours (SDSS~J081740.52+135134.5) and 
2.4~hours (SDSS~J120110.31+211758.5), with a full width at half maximum (FWHM) 
resolution of $\sim$8~km~s$^{-1}$. The data were reduced using standard reduction
procedures ({\it 6\/}). The H\,\textsc{i} column density was derived from fitting
a Voigt profile to the damped Ly~$\upalpha$ profile. The metallicity of the DLAs ([M/H]) is defined
by the ratio of metals to hydrogen compared to the solar abundance ({\it 38\/}): 
[M/H] = $\log N$(M) - $\log N$(H\,\textsc{i}) - $\log N$(M)$_\odot$ + $\log N$(H\,\textsc{i})$_\odot$.
It was measured from the column density of sulfur and silicon for the DLAs toward
SDSS~J081740.52+135134.5 and SDSS~J120110.31+211758.5, respectively. The properties
of the DLAs are listed in Table~S3, whereas the properties of the two background quasars
are tabulated in Table~S4.\\

\noindent\underline{Observational Properties}

The continuum flux density of each DLA target was determined by integrating the observed 
flux density over the full extent of the continuum emission feature in the continuum image. 
The quoted uncertainty is the standard deviation of the observed background noise in the 
continuum image from an annulus surrounding the detected emission with outer radius 
three times the size of the observed extent of the emission. The flux density and uncertainty 
for the velocity bins in the [C\,\textsc{ii}]~158$\upmu$m spectral cube were calculated in a 
similar manner. From these velocity profiles we determined the FWHM, the redshift and the 
integrated flux density of the [C\,\textsc{ii}]~158$\upmu$m line, which were converted 
to the total [C\,\textsc{ii}]~158$\upmu$m line luminosity, $L_{\rm [CII]}$, using standard 
procedures ({\it 39\/}). We used the fitter function in CASA to determine the 
physical sizes of both the continuum and the integrated flux density of the 
[C\,\textsc{ii}]~158$\upmu$m line. These measurements are tabulated in Table~S2.

To determine the velocity profile and axis used for creating the {\it p-v\/} diagrams, 
we generated images of the individual channel maps. These are shown in Figs.~S1 and S2. 
The direction of the axis was determined by fitting a straight line to the positions of 
maximum flux density in each velocity bin. This yielded position angles of $90 \pm 10$ degrees east of 
north and $130 \pm 15$ degrees west of north for ALMA~J081740.86+135138.2 and 
ALMA~J120110.26+211756.2, respectively. Uncertainties were determined through bootstrapping.
For ALMA J120110.26+211756.2, the individual channel maps show that at $\sim$ 65~km~s$^{-1}$, 
the emission shifts to the northwest by $ \sim 0.5''$. This indicates a more complicated velocity 
profile then a simple rotating disk, such as one arising from a merger or outflow.

In addition to continuum measurements of the galaxies associated with the DLAs, both quasars
were also detected in the 400~GHz continuum image. The continuum flux densities for the 
individual quasars are $0.54 \pm 0.08$ mJy and $13.15 \pm 0.08$ mJy, for SDSS~J081740.52+135134.5 
and SDSS~J120110.31+211758.5, respectively. SDSS~J120110.31+211758.5 was detected at 
high significance, allowing us to search for absorption features in the dust continuum of the quasar 
from both atoms and molecules. No absorption features were found. The 3$\sigma$ velocity-integrated
line opacity limit for [C\,\textsc{ii}]~158$\upmu$m absorption is $\sim 20$~km~s$^{-1}$, 
which corresponds to a C\,\textsc{ii} column density
of $N$(C\,\textsc{ii}) $< 2.8 \times 10^{18}$~cm$^{-2}$ ({\it 40\/}). This is in 
agreement with the estimated column density of C\,\textsc{ii} in the DLA 
based on the optical observations, $N$(C\,\textsc{ii}) $= 1.1 \pm 0.3 \times 10^{17}$~cm$^{-2}$. 
This column density estimate was determined from the Si\,\textsc{ii} column density 
measurement ({\it 6\/}), assuming a solar abundance ratio between carbon and 
silicon ({\it 38\/}).\\

\noindent\underline{Physical Parameters of the DLA Host Galaxies}

The SFR for the two galaxies is estimated from the continuum luminosity at a rest-frame 
wavelength of 158~$\upmu$m, as tabulated in Table~S2. To convert these luminosities to 
SFR estimates, we use the standard calibration obtained from nearby star-forming
galaxies ({\it 41\/}). Here, we note that our measured luminosities are at a rest 
wavelength of 158~$\upmu$m, and not 160~$\upmu$m, which, for the dust continuum 
models used in this paper, will overestimate the SFR by less than 5~\%. For galaxies 
with small SFRs, part of the 160~$\upmu$m emission arises from sources not associated 
with star formation ({\it 42\/}). However, the measured luminosities 
and predicted SFRs for the two galaxies are large enough that the contribution from 
non-stellar sources to the 160~$\upmu$m emission is likely to be negligible. The 
quoted uncertainties on the SFR are statistical; systematic errors due to the conversion 
are expected to result in an additional uncertainty by a factor of $\sim 2.5$ in SFR ({\it 41\/}). 

The continuum flux density estimate at a rest wavelength of 158~$\upmu$m can also
be used to estimate the total infrared (TIR) luminosity, which is defined as the total 
luminosity integrated over the interval between 8~$\upmu$m and 1000~$\upmu$m.
Because we have only a single point for the determination of the far-infrared spectral 
energy distribution, we apply a range of modified black body models, and fit these 
models to the measured 158~$\upmu$m continuum flux density. In particular, we 
assume the mid-infrared slope, $\alpha$, lies in the range $1.5 - 2.5$, 
the Raleigh-Jeans index, $\beta$, in the range $1.2 - 2.0$, and 
the dust temperature in the range $25 - 45$~K, consistent with other high redshift studies
({\it 24\/}). This results in a systematic uncertainty on the TIR by a factor of 2.5. 
The quoted uncertainty on the TIR luminosity includes both the uncertainty in the flux 
density measurement and this systematic uncertainty.

The dynamical mass of the galaxy that gives rise to the [C\,\textsc{ii}]~158$\upmu$m line
and the dust continuum emission for ALMA~J081740.86+135138.2 can be estimated from 
the velocity profile of the [C\,\textsc{ii}]~158$\upmu$m line. We estimate the dynamical mass 
with the equation: $M_{\rm dyn}/{\rm M}_\odot \approx 1.16 \times 10^5~v_{\rm circ}^2 D$, 
where $D$ is the physical diameter (in kpc) of the [C\,\textsc{ii}]~158$\upmu$m emission 
and $v_{\rm circ}$ is the circular velocity (in km~s$^{-1}$) ({\it 43\/}). The circular velocity is estimated 
by fitting an exponential disk model to the full spectral cube ({\it 44\/}). 
In this exponential disk model, we have fixed the inclination to 90 degrees 
(i.e., an edge-on disk) since the velocity resolution of the data is insufficient to 
accurately estimate the inclination. This yields a lower limit to the circular velocity and, 
hence, a lower limit to the dynamical mass. This lower limit is consistent with the 
circular velocity estimate obtained using $v_{\rm circ} = 0.75 \times {\rm FWHM_{[CII]}}$
({\it 43\/}). The resultant dynamical mass estimate is tabulated in Table~S2.

\newpage

\begin{figure}[!h]
\centering
\includegraphics[width=\textwidth]{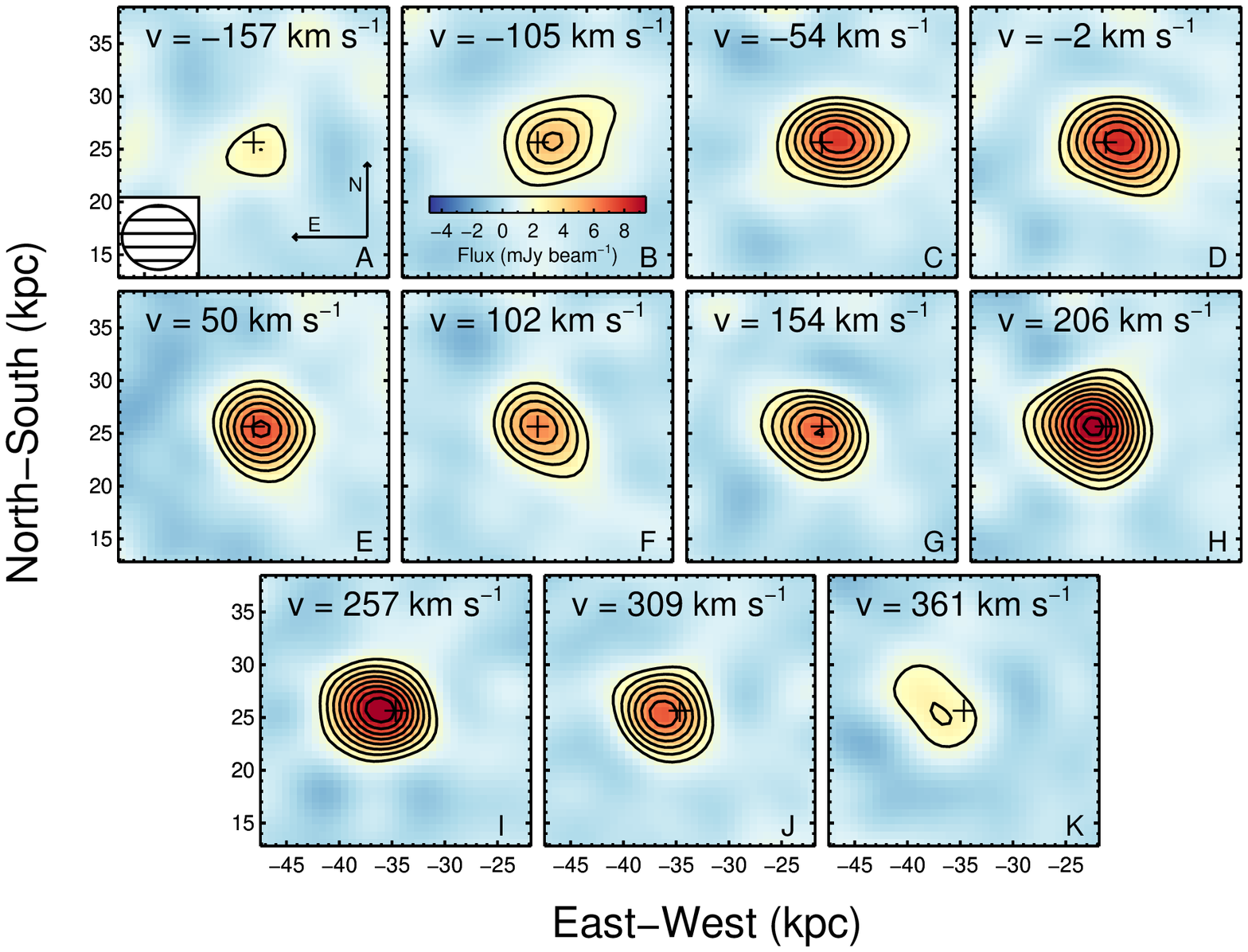}
\end{figure}
\noindent {\bf Fig. S1. Channel maps of the [C\,\textsc{ii}]~158$\upmu$m line emission
from ALMA~J081740.86+135138.2.} The plus symbol indicates the observed central position
of the [C\,\textsc{ii}] emission integrated over the full velocity profile (Fig.~1C). Axes give physical 
distances at the redshift of the DLA, in which the origin is set at the position of the DLA, as in
Fig~1. The outer black contour is drawn at 3$\sigma$ significance with subsequent contours 
increasing by $\sqrt{2}\sigma$. Velocities are relative to the DLA redshift ($z = 4.2584$). The lower
left inset in panel A is the size of the synthesized beam. The flux scale for each panel, given in
panel B, is the same for each panel.

\newpage

\begin{figure}[!h]
\centering
\includegraphics[width=\textwidth]{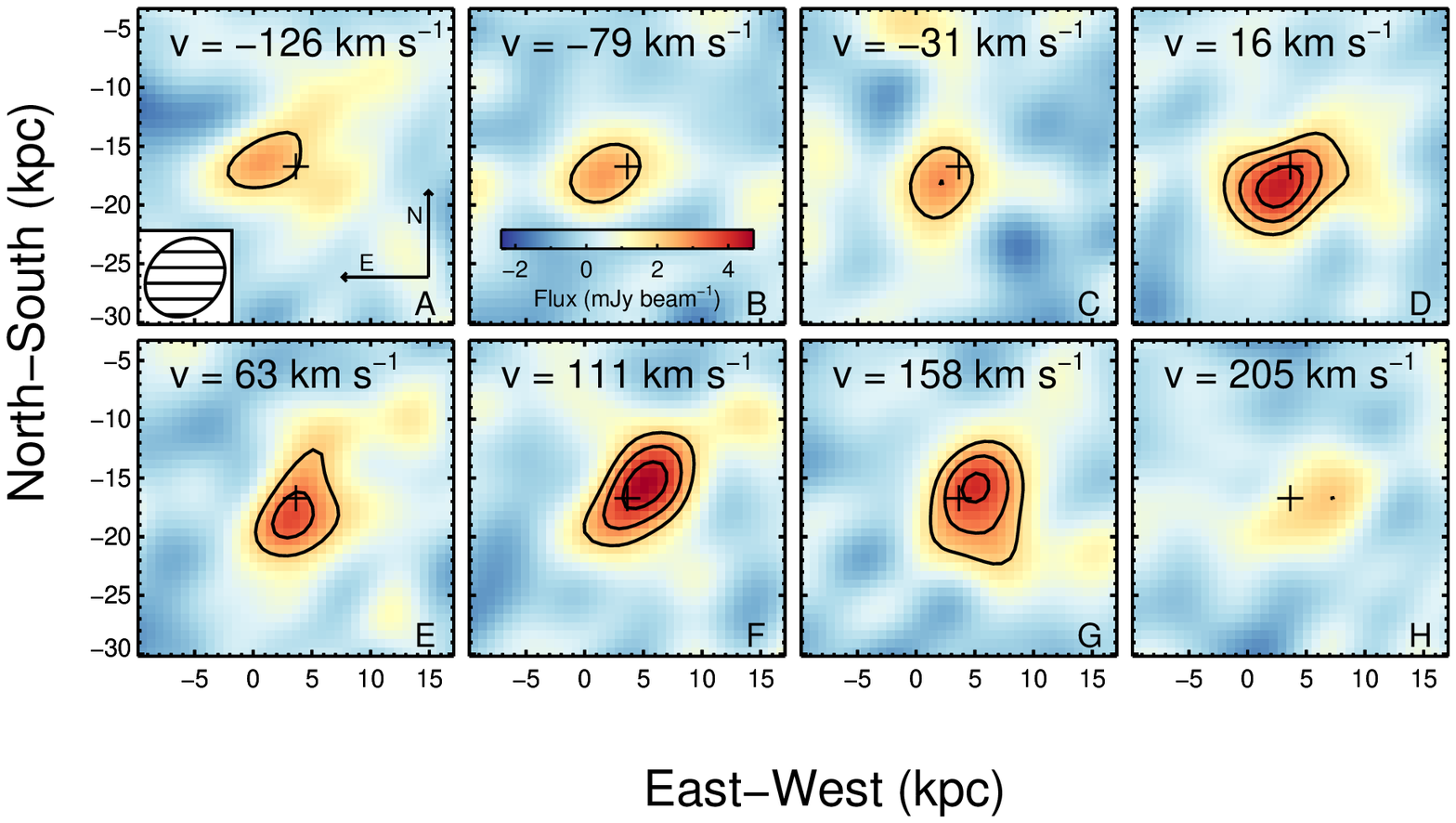}
\end{figure}
\noindent {\bf Fig. S2. Channel maps of the [C\,\textsc{ii}]~158$\upmu$m line emission from
ALMA J120110.26+211756.2.} The plus symbol indicates the observed central position
of the [C\,\textsc{ii}] emission integrated over the full velocity profile (Fig.~1D). Axes give physical 
distances at the redshift of the DLA, in which the origin is set at the position of the DLA, as in
Fig~1. The outer black contour is drawn at 3$\sigma$ significance with subsequent contours 
increasing by $\sqrt{2}\sigma$. Velocities are relative to the DLA redshift ($z = 3.7975$). The lower
left inset in panel A is the size of the synthesized beam. The flux scale for each panel, given in
panel B, is the same for each panel.

\newpage

\noindent {\bf Table S1. ALMA Observing Log.} Observing log of the ALMA observations
for the fields surrounding the quasars SDSS~J081740.52+135134.5 and 
SDSS~J120110.31+211758.5. The central frequency (CF), number of channels,
 and channel width are listed for each of the 4 individual sidebands.\\
\\
\begin{tabular}{llll}
& & SDSS~J081740.52+135134.5 & SDSS~J120110.31+211758.5\\
\hline
\hline
Date Observed & (UT) &2015 December 30 & 2015 December 29, 2016 January 1\\
On-source time & (s) & 2782 & 4596\\
Right Ascension & (J2000) & 08:17:40.53 & 12:01:10.31\\
Declination & (J2000) & +13:51:34.6 & +21:17:58.5\\
Number of antennas & & 39 & 44\\
Maximum Baseline & (m) & 263 & 263\\
Band 1 CF & (GHz) & 361.4 & 396.2\\
Band 2 CF & (GHz) & 363.3 & 394.5\\
Band 3 CF & (GHz) & 349.3 & 406.2\\
Band 4 CF & (GHz) & 351.1 & 407.9\\
Number of Channels & & 128 & 128\\
Channel width & (MHz) & 15.625 & 15.625\\
\hline
\end{tabular}

\newpage

\noindent {\bf Table S2. Measured properties for the DLA host galaxies.} 
The redshift ($z_{\rm [CII]}$), peak flux density ($S_{\rm [CII],peak}$), velocity integrated 
flux density ($\int S_{\rm [CII]} {\rm d}v$) and full width at half maximum (FWHM$_{\rm [CII]}$) are 
measurements from the ALMA spectral data cube, whereas the continuum flux density 
measurement ($S_{\rm cont}$) is from the continuum image. The sizes of the continuum 
emission ($A_{\rm cont}$) and [C\,\textsc{ii}] emission ($A_{\rm [CII]}$) are after deconvolution 
with the ALMA beam.\\
\\
\begin{tabular}{llll}
& & ALMA J081740.86+135138.2 & ALMA J120110.26+211756.2\\
\hline
\hline
Right Ascension & (J2000) & 08:17:40.86 & 12:01:10.26\\
Declination & (J2000) & +13:51:38.2 & +21:17:56.2\\
$z_{\rm [CII]}$ & & $4.2601 \pm 0.0001$ & $3.7978 \pm 0.0001$\\
$S_{\rm cont}$ & (mJy) & $1.19 \pm 0.11$ & $0.31 \pm 0.11$\\
$S_{\rm [CII],peak}$ & (mJy) & $13.1 \pm 1.9$ & $7.1 \pm 1.7$\\
FWHM$_{\rm [CII]}$ & (km~s$^{-1}$) & $460 \pm 50$ & $330 \pm 50$\\
$\int S_{\rm [CII]} {\rm d}v$ & (Jy~km~s$^{-1}$) & $5.4 \pm 0.6$ & $1.9 \pm 0.3$\\
$A_{\rm cont}$ & ($''$) & $(0.6 \pm 0.2) \times (0.3 \pm 0.3)$ & $(<1.4) \times (<0.4)^\ast$\\
$A_{\rm [CII]}$ & ($''$) & $(0.6 \pm 0.2) \times (0.3 \pm 0.2)$ & $(1.0 \pm 0.3) \times (0.4 \pm 0.3)$\\
$\log (L_{\rm TIR}/{\rm L}_\odot)$ & & $12.0 \pm 0.1 \pm 0.4$ & $11.4 \pm 0.2 \pm 0.4$\\
$\log (L_{\rm [CII]}/{\rm L}_\odot)$ & & $9.48 \pm 0.05$ & $8.94 \pm 0.07$\\
SFR & ($M_\odot$~yr$^{-1}$) & $110 \pm 10$ & $24 \pm 8$\\
$\log (M_{\rm dyn}/{\rm M}_\odot)$ & & $>10.8$ & ---$^\dagger$\\
\hline
\multicolumn{4}{l}{\small $^\ast$Unresolved.}\\
\multicolumn{4}{l}{\small $^\dagger$Not estimated.}\\
\end{tabular}

\newpage

\noindent {\bf Table S3. Measured properties of the DLAs.} The DLA redshift ($z_{\rm abs}$), 
the neutral hydrogen column density ($N$(H\,\textsc{i})) and metallicity ([M/H]) for
the DLAs are derived from the optical absorption spectra ({\it 6\/}). The elements
used for determining the metallicity are given in parentheses.\\
\\
\begin{tabular}{llll}
& & DLA~J081740.52+135134.5 & DLA~J120110.31+211758.5\\
\hline
\hline
Right Ascension & (J2000) & 08:17:40.52 & 12:01:10.31\\
Declination & (J2000) & +13:51:34.5 & +21:17:58.4\\
$z_{\rm abs}$ & & $4.2584 \pm 0.0001$ & $3.7975 \pm 0.0001$\\
$\log (N$(H\,\textsc{i})/cm$^{-2}$) & & $21.30 \pm 0.15$ & $21.35 \pm 0.15$\\
{[M/H]} & & $-1.15 \pm 0.15$ (S) & $-0.747 \pm 0.15$ (Si)\\
\hline
\end{tabular}

\newpage

\noindent {\bf Table S4. Measured properties of the Quasars.} The position of the
quasar, its redshift ($z_{\rm QSO}$), and photometry ($r$ and $i$ filters) are 
from the Sloan Digital Sky Survey (SDSS) ({\it 45\/}). The quasar
continuum flux densities ($S_{\rm cont}$) at the observed frequency, noted in 
parentheses, are from the ALMA continuum images.\\
\\
\begin{tabular}{llll}
& & SDSS~J081740.52+135134.5 & SDSS~J120110.31+211758.5\\
\hline
\hline
Right Ascension & (J2000) & 08:17:40.52 & 12:01:10.31\\
Declination & (J2000) & +13:51:34.5 & +21:17:58.4\\
$z_{\rm QSO}$ & & $4.398 \pm 0.001$ & $4.579 \pm 0.001$\\
$r$ & (AB) & $19.22 \pm 0.02$ & $20.38 \pm 0.03$\\
$i$ & (AB) & $19.87 \pm 0.02$ & $18.71 \pm 0.01$\\
$S_{\rm cont}$ & (mJy) & $0.55 \pm 0.09$ (350GHz) & $13.14 \pm 0.06$ (400GHz)\\
\hline
\end{tabular}

\end{document}